\documentclass[conference]{IEEEtran}
\IEEEoverridecommandlockouts
\usepackage{cite}
\usepackage{amsmath,amssymb,amsfonts}
\usepackage{algorithmic}
\usepackage{graphicx}
\usepackage{textcomp}
\usepackage{xcolor}
\usepackage{booktabs}
\def\BibTeX{{\rm B\kern-.05em{\sc i\kern-.025em b}\kern-.08em
    T\kern-.1667em\lower.7ex\hbox{E}\kern-.125emX}}
\begin{document}

\title{Reasoning Topology Matters: A Controlled Study of LLM-Based Cybersecurity Analysis \\}

\author{\IEEEauthorblockN{1\textsuperscript{st} Jiling Zhou\textsuperscript{\#,*}}
\IEEEauthorblockA{\textit{Department of Computing} \\
\textit{University of Turku}\\
Turku, Finland \\
ORCID: 0009-0006-7903-9355}
\textsuperscript{*}Corresponding author.
% ~\\

\and
\IEEEauthorblockN{2\textsuperscript{nd} Aisvarya Adeseye\textsuperscript{\#}}
\IEEEauthorblockA{\textit{Department of Computing} \\
\textit{University of Turku}\\
Turku, Finland \\
ORCID: 0009-0003-2401-3076}
\textsuperscript{\#}These authors contributed equally. 
% ~\\

\and
\IEEEauthorblockN{3\textsuperscript{rd} Antti Hakkala}
\IEEEauthorblockA{\textit{Department of Computing} \\
\textit{University of Turku}\\
Turku, Finland \\
ORCID: 0000-0002-0932-7814}
~\\

% \and
\and[\hfill\break\hfill]

\IEEEauthorblockN{4\textsuperscript{th} Seppo Virtanen}
\IEEEauthorblockA{\textit{Department of Computing} \\
\textit{University of Turku}\\
Turku, Finland \\
ORCID: 0000-0002-9487-3018}

\and
\IEEEauthorblockN{5\textsuperscript{th} Jouni Isoaho}
\IEEEauthorblockA{\textit{Department of Computing} \\
\textit{University of Turku}\\
Turku, Finland \\
ORCID: 0000-0002-5789-3992}

}

\maketitle

\begin{abstract}
Large Language Models (LLMs) are increasingly used in cybersecurity, where accurate analysis often requires multi-step and context-dependent reasoning over complex and heterogeneous data. However, existing prompting approaches typically focus on eliciting reasoning without explicitly considering how intermediate reasoning steps are structurally organized. We introduce \textit{Security Reasoning Topology}, which models reasoning through three representative structures: \textbf{Linear}, \textbf{Branching}, and \textbf{Graph}. To evaluate their effects, we conduct controlled experiments on three cybersecurity datasets covering MITRE ATT\&CK network traffic, cyber threat intelligence (CTI), and CVE vulnerability analysis. We evaluate multiple LLMs, including Llama 2 (7B, 13B, 70B), GPT-5.1, and Mistral Large 3, while keeping task inputs consistent and controlling reasoning structure through system-level prompting. Results show that reasoning topology substantially affects performance: Graph reasoning achieves the highest overall accuracy, improving over few-shot prompting by 9.8--12.2 percentage points across datasets, while Branching provides a strong intermediate solution. The results further show that the effect of reasoning topology remains consistent across model families and scales, highlighting reasoning topology as an important design factor for LLM-based cybersecurity analysis.
\end{abstract}


\section{Introduction}

Large Language Models (LLMs) are increasingly used in cybersecurity tasks such as threat detection, vulnerability analysis, cyber threat intelligence (CTI), and incident response \cite{zhang2025a,ferrag2024,xu2025a,jelodar2025}. In these settings, the quality of model reasoning can directly affect how security evidence is interpreted and how analytical conclusions are formed \cite{jin2024,atlam2025a,he2026}. Cybersecurity tasks, however, differ substantially in their reasoning requirements. Vulnerability analysis often follows sequential cause--effect relationships, CTI analysis may require considering multiple competing explanations, while attack-path analysis requires reasoning over interconnected behaviors and dependencies. Applying the same reasoning strategy across these structurally different tasks may therefore limit analytical performance.

Existing work has extensively studied prompting and reasoning methods such as Chain-of-Thought (CoT), Tree-of-Thought (ToT), and Graph-of-Thought (GoT) \cite{wei2022,yao2023,besta2024,yao2024}, while cybersecurity research has largely evaluated LLMs on individual security applications. However, these two directions are typically studied separately: reasoning methods are treated as independent prompting techniques, and limited attention has been given to the \emph{structural organization} of reasoning across cybersecurity tasks. This leaves an important question unresolved: how does the structure of intermediate reasoning itself influence cybersecurity analysis?

To investigate this question, we introduce \textit{Security Reasoning Topology}, a unified structural view of LLM reasoning. We characterize three representative topologies: \textbf{Linear}, which follows a sequential reasoning path; \textbf{Branching}, which explores multiple candidate paths; and \textbf{Graph}, which supports interconnected dependencies, information reuse, and iterative refinement. These structures are instantiated using CoT, ToT, and GoT, respectively. We evaluate them under a controlled setting across three cybersecurity datasets covering MITRE ATT\&CK network traffic, CTI, and CVE vulnerability analysis. To isolate the effect of reasoning topology, task-specific user inputs are kept consistent while system-level instructions control the reasoning structure across models.

\begin{figure*}[!t]
    \centering
    \includegraphics[width=0.85\textwidth]{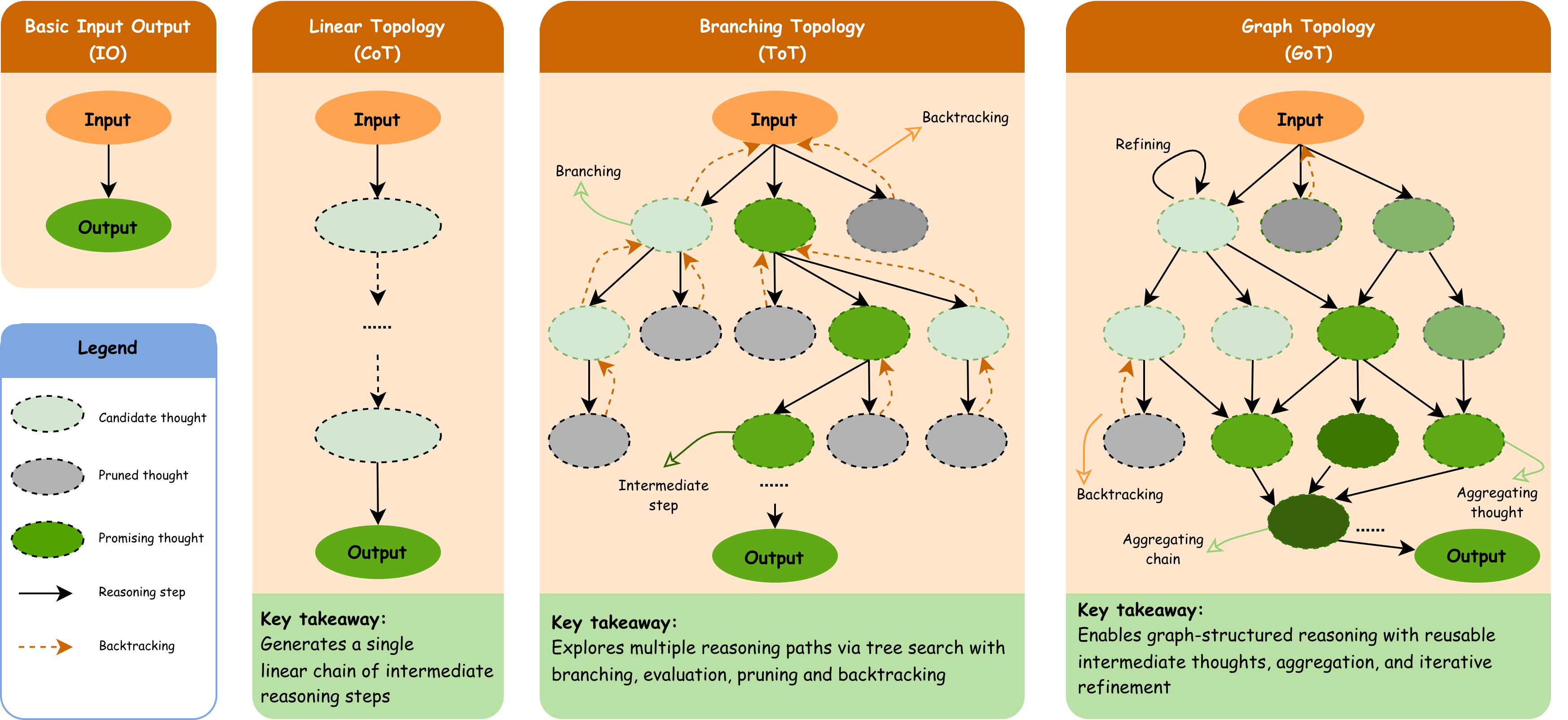}
    \caption{Reasoning paradigms interpreted as Linear, Branching, and Graph topologies, differing in how intermediate reasoning states
    are organized and connected.}
    \label{fig:reasoning_topologies}
\end{figure*}

Our study addresses two research questions: \textit{RQ1: How does reasoning topology affect LLM performance across cybersecurity analysis tasks?} and \textit{RQ2: How consistent are topology effects across model scales, and how does prompt organization affect performance under topology-based reasoning?} Experiments across multiple LLM families and model scales show that reasoning topology has a measurable impact on performance. Graph reasoning achieves the strongest overall accuracy, while Branching provides an intermediate solution between simpler Linear reasoning and more complex Graph reasoning.

The main contributions of this work are:

\begin{itemize}
    \item We introduce a \textbf{reasoning-topology perspective} for LLM-based cybersecurity analysis, abstracting representative reasoning paradigms into Linear, Branching, and Graph structures.
    
    \item We develop a \textbf{controlled evaluation framework} that separates system-level reasoning control from task-specific user inputs, enabling consistent comparison of reasoning topologies under identical task formulations.
    
    \item We conduct an empirical evaluation across heterogeneous cybersecurity datasets and multiple LLMs, demonstrating that topology effects remain consistent across model scales and that reasoning performance is influenced by prompt organization.
\end{itemize}

\section{Related Work}
\label{sec:related}

Research on LLM reasoning has progressively moved from sequential to more structured reasoning paradigms. Chain-of-Thought (CoT) elicits intermediate reasoning steps along a linear path \cite{wei2022,qiao2023}, while Tree-of-Thought (ToT) extends this process by exploring and evaluating multiple candidate paths \cite{yao2023}. Graph-of-Thought (GoT) further introduces graph-structured reasoning, enabling information aggregation, reuse, and refinement across interconnected reasoning states \cite{besta2024,yao2024}. Other approaches, such as self-consistency and Forest-of-Thought, improve robustness through sampling or combining multiple reasoning paths \cite{wang2023,bi2025}. Despite these advances, such methods are commonly studied as separate prompting or search paradigms rather than as comparable structural forms of reasoning.

In cybersecurity, LLMs have been investigated for vulnerability analysis, cyber threat intelligence (CTI), and MITRE ATT\&CK-based security analysis. Prior work has explored vulnerability understanding and classification \cite{zhang2025a,karras2025a}, threat-report analysis and technique extraction \cite{perrina2023a,büchel2025}, and ATT\&CK mapping and attack reasoning \cite{fayyazi2024,jin2024}. These tasks exhibit different reasoning characteristics: vulnerability analysis often involves sequential causal interpretation, CTI may require reasoning over ambiguous or competing evidence, and ATT\&CK analysis can involve dependencies across multiple attack stages. However, existing cybersecurity studies largely focus on task-specific performance, while reasoning research focuses on individual reasoning methods. Consequently, systematic comparisons of how different reasoning structures behave across heterogeneous cybersecurity tasks remain limited. Our work bridges these directions by modeling Linear, Branching, and Graph reasoning under a unified topology perspective and evaluating them across multiple cybersecurity datasets and LLMs.

\section{Security Reasoning Topology}
\label{sec:topology}

Cybersecurity tasks differ in how intermediate evidence must be organized during reasoning. As discussed in Section~\ref{sec:related}, vulnerability analysis often involves sequential cause--effect interpretation, CTI analysis may involve multiple plausible hypotheses, and ATT\&CK-based analysis can require reasoning over interconnected attack behaviors. Motivated by these differences, we introduce \textit{Security Reasoning Topology}, which models LLM reasoning as an explicit structural process rather than an implicit sequence of generated steps.

For a cybersecurity task $\mathbf{T}$, we represent its reasoning topology as
\begin{equation}
    \mathbf{R} = (\mathbf{V}, \mathbf{E}),
\end{equation}
where $\mathbf{V}$ denotes intermediate reasoning states and $\mathbf{E}$ represents dependencies or transitions among them. Different reasoning strategies can therefore be characterized by how these states are organized and connected.

We consider three representative topologies, illustrated in Fig.~\ref{fig:reasoning_topologies}. \textbf{Linear reasoning} organizes states along a sequential path, where each step primarily depends on its predecessor. \textbf{Branching reasoning} generates and evaluates multiple candidate paths, enabling comparison among alternative hypotheses. \textbf{Graph reasoning} allows states to be interconnected, reused, merged, and refined, supporting reasoning over more complex dependencies.

These topologies provide a common structural abstraction for representative LLM reasoning paradigms. In our implementation, Linear, Branching, and Graph reasoning are instantiated using Chain-of-Thought (CoT), Tree-of-Thought (ToT), and Graph-of-Thought (GoT), respectively \cite{wei2022,yao2023,besta2024}. Rather than assuming a fixed mapping between cybersecurity tasks and optimal topologies, we empirically evaluate how reasoning-structure complexity affects performance across heterogeneous security tasks and whether these effects remain
consistent across model scales.

\section{Experimental Setup}
\label{sec:methodology}

\subsection{Datasets}

We evaluate reasoning topologies on three publicly available cybersecurity datasets covering heterogeneous data and task structures. The \textbf{ATT\&CK dataset} contains structured network traffic records annotated with MITRE ATT\&CK tactic labels, including protocol, service, connection state, packet and byte counts, ports, and timestamps \cite{elam2025}. The \textbf{CTI dataset} contains 1,100 threat reports with threat categories and indicators of compromise, providing unstructured textual inputs for threat analysis \cite{zotero-item-1483}. The \textbf{CVE dataset} contains structured vulnerability records with identifiers, categories, and technical descriptions \cite{zotero-item-cve}. Duplicate records are removed and samples are balanced where necessary to reduce dataset bias.

\subsection{Compared Methods}

We compare three topology-based reasoning methods with three non-topological baselines. The topology-based methods instantiate \textbf{Linear}, \textbf{Branching}, and \textbf{Graph} reasoning using CoT, ToT, and GoT, respectively. The baselines capture three complementary factors unrelated to explicit reasoning topology: \textbf{few-shot prompting} provides labeled demonstrations before prediction \cite{cheng2024}; \textbf{structured JSON prompting} constrains the output to a predefined schema \cite{agarwal2025a}; and \textbf{RAG} augments the model input with retrieved cybersecurity context such as ATT\&CK descriptions, CTI evidence, or CVE information \cite{fan2024}. This design allows us to compare explicit reasoning structure with improvements arising from demonstrations, output constraints, and external knowledge.

\subsection{Controlled Prompt Design and Models}

To control topology implementation, we separate reasoning instructions from task-specific inputs. The \textbf{system prompt} specifies how intermediate reasoning states should be organized for Linear, Branching, or Graph reasoning, while the \textbf{user prompt} contains the task instruction and security input. Within the topology-based conditions, the task formulation is kept fixed while only the system-level reasoning structure changes, allowing the effect of topology to be compared under consistent inputs.

The topology implementations are controlled at the prompt level rather than through a separate external orchestration framework; the system prompt specifies the structural instructions appropriate
to each reasoning topology, including branching, dependency, and
refinement when applicable.

We evaluate Llama 2 models with 7, 13, and 70 billion parameters, GPT-5.1, and Mistral Large 3 to examine whether topology effects persist across model families and scales. Temperature and top-$p$ are fixed at $0.2$ and $0.9$, respectively. The maximum generation length is 1024 tokens, extended to 2048 for ToT and GoT to accommodate expanded reasoning structures. Temperature $0.2$ was selected through a pilot ablation over $\{0.0,0.2,0.5\}$. For each task instance, model outputs are recorded under the same evaluation protocol.

\subsection{Evaluation Metrics}

We use \textbf{Accuracy} as the primary evaluation metric, measuring the proportion of predictions that match the ground-truth task label. Each model--method--dataset configuration is evaluated over five independent runs. Aggregate results are reported across the five evaluated models and five runs. Paired bootstrap tests are used to evaluate the statistical significance of performance differences between methods.

\section{Results and Discussion}
\label{sec:results}

\subsection{Overall Performance}

Table~\ref{tab:main_results} summarizes aggregate accuracy across the five evaluated models and five independent runs on the three cybersecurity datasets. All topology-based methods improve over direct few-shot prompting, while the magnitude of improvement generally increases with reasoning-structure complexity. GoT achieves the highest mean accuracy on all three datasets, reaching 85.0\% on ATT\&CK and CTI and 82.0\% on CVE. Relative to few-shot prompting, these correspond to improvements of 9.8, 12.2, and 11.8 percentage points, respectively.

\begin{table}[!t]
\centering
\caption{Aggregate accuracy across five evaluated models and five independent runs on the three cybersecurity datasets. Values are reported as mean $\pm$ standard deviation.}
\label{tab:main_results}
\small
\setlength{\tabcolsep}{4pt}
\begin{tabular}{lccc}
\toprule
\textbf{Method} & \textbf{ATT\&CK} & \textbf{CTI} & \textbf{CVE} \\
\midrule
Few-shot & $75.2\pm1.8$ & $72.8\pm2.1$ & $70.2\pm2.0$ \\
JSON     & $77.2\pm1.7$ & $76.2\pm1.9$ & $73.4\pm1.8$ \\
RAG      & $80.6\pm1.6$ & $80.8\pm1.7$ & $76.8\pm1.7$ \\
CoT      & $78.2\pm1.6$ & $78.2\pm1.8$ & $75.2\pm1.7$ \\
ToT      & $82.2\pm1.5$ & $82.2\pm1.6$ & $79.2\pm1.6$ \\
\textbf{GoT}
         & $\mathbf{85.0\pm1.4}$
         & $\mathbf{85.0\pm1.5}$
         & $\mathbf{82.0\pm1.5}$ \\
\bottomrule
\end{tabular}
\end{table}

RAG is the strongest non-topological baseline, achieving 80.6\%, 80.8\%, and 76.8\% accuracy on ATT\&CK, CTI, and CVE, respectively. Nevertheless, GoT remains consistently stronger, indicating that explicitly organizing intermediate reasoning can provide benefits beyond supplying additional external knowledge. ToT provides a strong intermediate topology, while CoT produces smaller but consistent gains over few-shot prompting.

Paired bootstrap tests further support these differences. GoT significantly outperforms few-shot prompting on all three datasets ($p<0.001$) and also improves over RAG ($p<0.05$).

\subsection{Model-Scale Effects}

% The effect of reasoning topology remains consistent across the evaluated model families and scales. For Llama 2 7B, 13B, and 70B, as well as GPT-5.1 and Mistral Large 3, GoT achieves the highest accuracy within each model group, followed by ToT and CoT. Within the Llama 2 family, absolute performance also increases with model size while the relative ordering of the three reasoning topologies remains stable. This suggests that the observed topology effect is not restricted to a single model or parameter scale and that models with different capacities can benefit from explicitly structured reasoning.

Table~\ref{tab:model_scale} shows that GoT consistently achieves the highest average accuracy within every evaluated model group, followed by ToT and CoT. Within the Llama 2 family, absolute performance increases from 7B to 70B while the ordering of the three topologies remains stable. The same ordering is observed for GPT-5.1 and Mistral Large 3, indicating that the topology effect is consistent across different model families and scales.

\begin{table}[!t]
\centering
\caption{Average accuracy across the three datasets for different models and reasoning topologies.}
\label{tab:model_scale}
\small
\setlength{\tabcolsep}{4pt}
\begin{tabular}{lccc}
\toprule
\textbf{Model} & \textbf{CoT} & \textbf{ToT} & \textbf{GoT} \\
\midrule
Llama 2 7B   & 65.7 & 69.7 & \textbf{72.7} \\
Llama 2 13B  & 71.3 & 75.3 & \textbf{78.3} \\
Llama 2 70B  & 79.3 & 83.3 & \textbf{86.3} \\
GPT-5.1      & 86.3 & 90.3 & \textbf{93.3} \\
Mistral Large 3 & 83.3 & 87.3 & \textbf{89.3} \\
\bottomrule
\end{tabular}
\end{table}

\subsection{Prompt Organization Ablation}
We further examine whether system--user prompt separation affects the performance of topology-based reasoning. For each topology, we compare three settings: using only a user prompt, merging topology and task instructions into a single user prompt, and separating topology control into the system prompt while keeping task-specific inputs in the user prompt.

The separated design consistently achieves the highest accuracy.
Averaged across the three datasets, separating system and user prompts improves accuracy over the user-only setting by 1.9 percentage points for CoT, 2.9 points for ToT, and 3.7 points for GoT. The larger gain for GoT suggests that more complex reasoning structures particularly benefit from explicit system-level topology control. These results support system--user prompt decomposition as a controlled implementation mechanism for reasoning topology.

\subsection{Discussion}

Overall, the results show that reasoning topology is a consistent factor in LLM-based cybersecurity analysis. Linear reasoning provides a simple structured approach, Branching improves performance through alternative-path exploration, and Graph reasoning achieves the strongest overall results across the evaluated tasks and models.
At the same time, the prompt ablation shows that reliable topology implementation depends on how structural instructions are separated from task-specific inputs. Together, these findings support reasoning topology as a useful abstraction for systematically studying and controlling LLM reasoning in cybersecurity.

Two limitations are particularly relevant. First, the current evaluation focuses on analytical performance and does not explicitly quantify token consumption or end-to-end inference latency; a fuller analysis of accuracy--cost trade-offs is left for future work. Second, we do not systematically categorize topology-specific failure modes. Possible failure modes include ineffective branch selection in Branching reasoning and inconsistent or hallucinated dependencies in Graph reasoning, which warrant further investigation.

\section{Conclusion}
\label{sec:conclusion}

This work investigates reasoning topology as a structural factor in LLM-based cybersecurity analysis. We formulate Linear, Branching, and Graph reasoning as three representative topologies and instantiate them using CoT, ToT, and GoT under a controlled prompt design. Experiments across three cybersecurity datasets and multiple LLMs show that reasoning topology consistently affects analytical performance, with Graph reasoning achieving the highest overall accuracy and Branching providing a strong intermediate approach. The results also show that these effects persist across model scales and that separating system-level topology instructions from task-specific user inputs improves performance under topology-based reasoning. Overall, the findings suggest that reasoning structure should be treated as an explicit design dimension when developing LLM-based cybersecurity analysis systems. Future work will investigate adaptive topology selection and extend the evaluation to broader security scenarios.

\section*{Acknowledgment}
This project has received funding from the European Union’s Horizon Europe research and innovation programme under the Marie Skłodowska-Curie grant agreement No 101177564 — HAIF.

%\section*{References}
\bibliographystyle{IEEEtran}
\bibliography{Reasoning_topology}

\end{document}